\documentclass[preprint,amsmath,amssymb,amsfonts,aps,superscriptaddress,nofootinbib]{revtex4-1}
\usepackage[pdftex]{graphicx}
\usepackage{epsfig}
\usepackage{subfigure,multirow,makecell}
\usepackage{ifsym}
\usepackage{MnSymbol}
\usepackage[usenames,dvipsnames]{xcolor}
\usepackage[utf8]{inputenc}
\usepackage{graphicx}
\def\be{\begin{equation}}
\def\ee{\end{equation}}
\def\ba{\begin{eqnarray}}\def\bea{\begin{eqnarray}}
\def\ea{\end{eqnarray}}  \def\eea{\end{eqnarray}}

\def\nn{\nonumber}

\def\l{\lambda}
\def\O{\Omega}\def\o{\omega}
\def\th{\theta}

\def\[{\left[}
\def\]{\right]}
\def\({\left(}
\def\){\right)}

\def\<{\langle}
\def\>{\rangle}

   \def\sp2n{Sp(2N)}

\def\2F1{\,_2{\rm F}_1}

\begin{document}
% Use the \preprint{} command to place your local institutional report
\title{Superradiant stability of five and six-dimensional extremal Reissner-Nordstrom black holes}

% repeat the \author .. \affiliation  etc. as needed
% \email, \thanks, \homepage, \altaffiliation all apply to the current
% author.
% \affiliation command applies to all authors since the last
% \affiliation command. The \affiliation command should follow the
% other information
% \affiliation can be followed by \email, \homepage, \thanks as well.
%\author{}
%\email[]{Your e-mail address}
%\homepage[]{Your web page}
%\thanks{}
%\altaffiliation{}
%\affiliation{}
\author{Jia-Hui Huang}
\email{huangjh@m.scnu.edu.cn}
\affiliation{Guangdong Provincial Key Laboratory of Quantum Engineering and Quantum Materials,
School of Physics and Telecommunication Engineering,
South China Normal University, Guangzhou 510006, China}
\affiliation{Guangdong Provincial Key Laboratory of Nuclear Science, Institute of quantum matter,
South China Normal University, Guangzhou 510006, China}
\affiliation{Guangdong-Hong Kong Joint Laboratory of Quantum Matter, Southern Nuclear Science Computing Center, South China Normal University, Guangzhou 510006, China}
\author{Tian-Tian Cao}
\author{Mu-Zi Zhang}
\affiliation{Guangdong Provincial Key Laboratory of Nuclear Science, Institute of quantum matter,
South China Normal University, Guangzhou 510006, China}
\affiliation{Guangdong-Hong Kong Joint Laboratory of Quantum Matter, Southern Nuclear Science Computing Center, South China Normal University, Guangzhou 510006, China}

%\date{\today}

\begin{abstract}
The superradiant stability of five and six-dimensional extremal Reissner-Nordstrom black holes under charged massive scalar perturbation is studied. In each case, it is analytically proved that the effective potential experienced by the scalar perturbation has only one maximum outside the black hole horizon and no potential well exists for the superradiance modes. So the five and six-dimensional extremal Reissner-Nordstrom black holes are superradiantly stable. In the proof, we develop a new method which is based on the Descartes' rule of signs for the polynomial equations. Our results generalize the previous study that four-dimensional extremal Reissner-Nordstrom black hole is superradiantly stable under charged massive scalar perturbation.
\end{abstract}

\maketitle

%\tableofcontents

\section{Introduction}
Superradiance is an interesting phenomenon in black hole physics \cite{Manogue1988,Greiner1985,Cardoso2004,Brito:2015oca,Brito:2014wla}. When a charged bosonic wave is impinging upon a charged rotating black hole, the wave is amplified by the black hole if the wave frequency $\omega$ obeys
  \begin{equation}\label{superRe}
   \omega < n\Omega_H  + e\Phi_H,
  \end{equation}
where $e$ and $n$ are the charge and azimuthal number of the bosonic wave mode, $\Omega_H$ is the angular velocity of the black hole horizon and $\Phi_H$ is the electromagnetic potential of the black hole horizon. This superradiant scattering was studied long time ago \cite{P1969,Ch1970,M1972,Ya1971,Bardeen1972,Bekenstein1973,Damour:1976kh}, and has broad applications in various areas of physics(for a comprehensive review, see\cite{Brito:2015oca}).

If there is a mirror mechanism that makes the amplified wave be scattered back and forth, it will lead to the superradiant instability of the background black hole geometry \cite{PTbomb,Cardoso:2004nk,Herdeiro:2013pia,Degollado:2013bha}. Superradiant (in)stability of various kinds of black holes has been studied extensively in the literature.
The superradiant (in)stability of rotating Kerr black holes under massive scalar perturbation has been studied in \cite{Huang:2019xbu,Strafuss:2004qc,Konoplya:2006br,Cardoso:2011xi,Dolan:2012yt,Hod:2012zza,Hod:2014pza,Aliev:2014aba,Hod:2016iri,Degollado:2018ypf}. Superradiant instability of a Kerr black hole that is perturbed by a massive vector field is also discussed in \cite{East:2017ovw,East:2017mrj}.
Rotating or charged black holes with asymptotically curved space are proved to be superradiantly unstable because the curved backgrounds provide  natural mirror-like boundary conditions
\cite{Cardoso:2004hs,Cardoso:2013pza,Zhang:2014kna,Delice:2015zga,Aliev:2015wla,Wang:2015fgp,
Ferreira:2017tnc,Wang:2014eha,Bosch:2016vcp,Huang:2016zoz,Gonzalez:2017shu,Zhu:2014sya}.

Among the study of superradiant (in)stability of black holes, an interesting result is that the  four-dimensional extremal and non-extremal Reissner-Nordstrom(RN) black holes have been proved to be superradiantly stable against charged massive scalar perturbation in the full parameter space of the black-hole-scalar-perturbation system\cite{Hod:2013eea,Huang:2015jza,Hod:2015hza,DiMenza:2014vpa}. The argument is that the two conditions for superradiant instability, (1) existence of a trapping potential well outside the black hole horizon and (2) superradiant amplification of the trapped modes, cannot be satisfied simultaneously \cite{Hod:2013eea,Hod:2015hza}.

In this paper, the study of superradiant stability of four-dimensional RN black holes will be generalized to higher dimensional. As a first step, we will analytically study the superradiant stability of five and six-dimensional extremal RN black holes under charged massive scalar perturbation. In Section II, we give a general description of the model we are interested in. In Section III and IV, we provide the proof for five and six-dimensional extremal RN black hole cases respectively. The last Section is devoted to conclusion and discussion.

\section{D-dimensional RN black holes and Klein-Gordon equation}
In this section we will give a description of the $D$-dimensional RN black hole and the Klein-Gordon equation for the charged massive scalar perturbation. The metric of the $D$-dimensional RN black hole \cite{Myers:1986un,Destounis:2019hca} is
\bea
ds^2=-f(r)dt^2+\frac{dr^2}{f(r)}+r^2d\O_{D-2}^2.
\eea
$d\O_{D-2}^2$ is the common line element of a $(D-2)$-dimensional unit sphere $ S^{D-2}$
\bea
d\O_{D-2}^2=d\th_{D-2}^2+\sum^{D-3}_{i=1} \prod_{j=i+1}^{D-2}\sin^2(\th_{j})d\th_i^2,
\eea
 where the ranges of the angular coordinates are  $\th_i\in [0,\pi](i=2,..,D-2), \th_1\in [0,2\pi]$.
$f(r)$ reads
\bea
f(r)=1-\frac{2m}{r^{D-3}}+\frac{q^2}{r^{2(D-3)}},
\eea
where the parameters $m$ and $q^2$ are related with the ADM mass $M$ and electric charge $Q$ of the RN black hole,
\bea
m=\frac{8\pi}{(D-2)Vol(S^{D-2})}M,~~ q= \frac{8\pi}{\sqrt{2(D-2)(D-3)}Vol(S^{D-2})}Q.
\eea
In the above equation, $Vol(S^{D-2})=2\pi^{\frac{D-1}{2}}/\Gamma(\frac{D-1}{2})$ is the volume of unit $(D-2)$-sphere. The inner and outer horizons of the RN black hole  are
$
  r_\pm=(m\pm\sqrt{m^2-q^2})^{1/(D-3)}.
$
For extremal RN black holes, the inner and outer horizons become one horizon
\bea
r_h=m^{1/(D-3)}.
\eea

 The electromagnetic field outside the black hole horizon is described
 by the following 1-form  vector potential
 \bea
 A=-\sqrt{\frac{D-2}{2(D-3)}}\frac{q}{r^{D-3}} dt=-c_D\frac{q}{r^{D-3}} dt.
 \eea

The equation of motion for a charged massive scalar perturbation in the RN black hole background is governed by
the following covariant Klein-Gordon equation
\bea
(D_\nu D^\nu-\mu^2)\phi=0,
\eea
where $D_\nu=\nabla_\nu-ie A_\nu$ is the covariant derivative and $\mu,~e$ are the mass and charge of the scalar field respectively. The solution with definite angular frequency for the above Klein-Gordon equation can be decomposed as
\bea
\phi(t,r,\th_i)=e^{-i\o t}R(r)
\Theta(\th_i).
\eea
The angular eigenfunctions $\Theta(\th_i)$  are $(D-2)$-dimensional scalar spherical harmonics and the eigenvalues are given by $-l(l+D-3), (l=0,1,2,..)$\cite{Chodos:1983zi,Higuchi:1986wu,Rubin1984,Achour:2015zpa,Lindblom:2017maa}.
The radial equation is
\bea\label{eq-radial}
\Delta\frac{d}{dr}(\Delta\frac{d R}{dr})+U R=0,
\eea
where $\Delta=r^{D-2}f(r)$ and
\bea
U=(\o+e A_t)^2 r^{2(D-2)}-l(l+D-3) r^{D-4}\Delta-\mu^2 r^{D-2}\Delta.
\eea

Define the tortoise coordinate $y$ by $dy=\frac{r^{D-2}}{\Delta}dr$ and a new radial function $\tilde{R}=r^{\frac{D-2}{2}}R$, then
the radial equation \eqref{eq-radial} can be rewritten as
\bea\label{tor-eq}
\frac{d^2\tilde{R}}{dy^2}+\tilde{U} \tilde{R}=0,
\eea
where
\bea
\tilde{U}=\frac{U}{r^{2(D-2)}}-\frac{(D-2)f(r)[(D-4)f(r)+ 2 r f'(r)]}{4r^2}.
\eea
The asymptotic behaviors of $\tilde{U}$ at the spatial infinity and outer horizon are
\bea
\lim_{r\rightarrow +\infty}\tilde{U}= \o^2-\mu^2,~~
\lim_{r\rightarrow r_+} \tilde{U}= (\o-c_D\frac{e q}{r_+^{D-3}})^2=(\o-e\Phi_h)^2,
\eea
where $\Phi_h$ is the electric potential of the outer horizon of the black hole.

The physical boundary conditions that we need are ingoing wave at the horizon ($y\to -\infty$) and bound states (exponentially decaying modes) at spatial infinity ($ y\to +\infty $). Then the asymptotic solutions of the equation \eqref{tor-eq} are as follows
\bea
r\to +\infty (y\to +\infty ),~\tilde{R}\sim {{e}^{-\sqrt{{{\mu }^{2}}-{{\omega }^{2}}}{y}}};\\
r\to {{r}_{+}}(y\to -\infty ),~\tilde{R}\sim {{e}^{-i(\omega -e\Phi_{h}){y}}}.
\eea
The exponentially decaying modes (bound state condition) require the following inequality
\bea\label{boundstate}
\omega^2<\mu^2.
\eea

Next we define a new radial function $\psi=\Delta^{1/2} R$, then the radial equation \eqref{eq-radial} can be written as a  Schrodinger-like equation
\bea
\frac{d^2\psi}{dr^2}+(\o^2-V)\psi=0,
\eea
where $V$ is the effective potential.

In the extremal case, the explicit expression for the effective potential $V$ is
\bea
V=\o^2+\frac{B}{A},
\eea
where $A$ and $B$ are
\bea\label{V}
A&=&4r^{2}(r^{2D-6}-2 m r^{D-3}+m^2)^2=4r^2(r^{D-3}-m)^4,\\
B&=&4(\mu^2-\o^2)r^{4D-10}+(2l+D-2)(2l+D-4)r^{4D-12}-8(m\mu^2-c_D e m \o)r^{3D-7}\nn\\
&-&4m(2\l_l+(D-4)(D-2))r^{3D-9}+4m^2(\mu^2-c_D^2 e^2)r^{2D-4}\nn\\
&+&2m^2(2 \l_l+3(D-4)(D-2))r^{2D-6}-4m^3(D-4)(D-2)r^{D-3}\nn\\
&+&m^4(D-4)(D-2),
\eea
and $\l_l=l(l+D-3)$.

The asymptotic behaviors of this effective potential $V$ at the horizon and spatial infinity are
\bea
V\rightarrow -\infty ,~~~~~r\rightarrow r_h;\\
V\rightarrow \mu^2 ,~~~~~r\rightarrow +\infty.
\eea
 At the spatial infinity, the asymptotic behavior of the derivative of the effective potential, $V'(r)$, is
 \bea\label{asymp}
V'(r)\rightarrow \left\{
   \begin{array}{ll}
     \frac{-(D-2)(D-4)-4\l_l-8m(\mu^2+c_D e \o-2\o^2)}{2r^3}, & \hbox{$D=5$;} \\
     \frac{-(D-2)(D-4)-4\l_l}{2r^3}, & \hbox{$ D\geqslant 6$.}
   \end{array}
 \right.
 \eea
The superradiance condition in the extremal case is
\bea\label{sup-con-extr}
\o<e\Phi_h=e c_D\frac{q}{r_h^{D-3}}=c_D e=\sqrt{\frac{D-2}{2(D-3)}} e.
\eea
Together with the bound state condition \eqref{boundstate}, we can prove $V'(r)<0$ at spatial infinity when $D=5$. It is also obvious that $V'(r)<0$ at spatial infinity when $D\geqslant 6$.
This means that there is no potential well when $ r\to +\infty$ and there is at least one extreme for the
effective potential $V(r)$ outside the horizon.

In the following sections, we will prove that there is indeed only one extreme outside the event horizon $r_h$ for the effective potential in each case of $D=5,6$  extremal RN black holes,  no potential well exists outside the event horizon for the superradiance modes. So the $D=5,6$ extremal RN black holes are superradiantly stable under massive scalar perturbation.

It is worth noting that the important mathematical theorem we will use in the proof is the\textit{ Descartes' rule of signs} which asserts that the number of positive roots of a polynomial equation with real coefficients is at most the number of sign changes in the sequence of the polynomial's coefficients.

\section{D=5 extremal RN black holes}
For a $D$=5 extremal RN black hole, the event horizon is located at $r^{(5)}_h=\sqrt{m}$.
The explicit expression of superradiance condition \eqref{sup-con-extr} is
\bea\label{sup-con-5D}
\o<\frac{\sqrt{3}}{2} e\approx 0.87 e.
\eea
In order to prove there is only one extreme outside the event horizon for the effective potential $V_5(r)$ in the D=5 case, we will consider the derivative of $V_5(r)$, i.e. $V_5'(r)$, and prove that there is only one real root for $V_5'(r)=0$ when $r>r^{(5)}_h$. The key result in the proof is summarized in Table I.
\begin{table}
\caption{Possible signs of $\{a_5,a_4,a_3,a_2,a_1,a_0\}$ in different intervals of $t$.}
\centering
\renewcommand{\multirowsetup}{\centering}
\begin{tabular}{|c|p{1.1cm}<{\centering}|p{1.1cm}<{\centering}|p{1.1cm}<{\centering}|p{1.1cm}<{\centering}
|p{1.1cm}<{\centering}|p{1.1cm}<{\centering}|}
\hline
% after \\: \hline or \cline{col1-col2} \cline{col3-col4} ...
$t$&$a_5$ & $a_4$ & $a_3$ & $a_2$ & $a_1$&$a_0$\\
\hline
(0.61,~0.87)&- & - & - & - & -& +\\
\hline
(0.41,~0.61)&- & - & - & - & +& +\\
\hline
\multirow{2}*{(0.25,~0.41)}&\multirow{2}*{-} & \multirow{2}*{-} & \multirow{2}*{-} & - & \multirow{2}*{+}& \multirow{2}*{+}\\
\cline{5-5}
& & & &+ & &\\
\hline
\multirow{3}*{(0.12,~0.25)}&\multirow{3}*{-} &\multirow{3}*{ -} & - & - & \multirow{3}*{+}& \multirow{3}*{+}\\
\cline{4-5}
& & &- &+ & &\\
\cline{4-5}
& & &+ &+ & &\\
\hline
\multirow{4}*{(0,~~~0.12)}&\multirow{4}*{-} & - & - & - & \multirow{4}*{+}&\multirow{4}*{+} \\
\cline{3-5}
& & -&- &+ & &\\
\cline{3-5}
& &- &+ &+ & &\\
\cline{3-5}
& & +&+ &+ & &\\
\hline
\end{tabular}\label{table}
\end{table}

 From the general expression of the effective potential \eqref{V}, we can calculate that the denominator of  $V_5'(r)$ is $2 r^3(r^2-m)^5$. The numerator of $V_5'(r)$ is
\bea\nn
n_5&=&3 m^5 - 15 m^4 r^2 + 2 m^3 r^4 (15 - 2l(l+2)) +
 2 m^2 r^6 (-15 + 3 e^2 m - 4 m \mu^2 + 2l(l+2)) \\&+&
 m r^8 (15 + 6 e^2 m + 16 m \mu^2 - 12 \sqrt{3} e m \o + 4l(l+2))\nn \\
 &-& r^{10} (3 + 8 m \mu^2 + 4 m (\sqrt{3} e - 4 \o) \o + 4 l(l+2)).
\eea
As mentioned before, we want to consider the real roots of $V_5'(r)=0$ when $r>r_h^{(5)}$. It is equivalent to considering the real roots of $n_5(r)=0$ when $r>r_h^{(5)}$.

Now we make a change for the variable $r$ and let $z=r^2-m$, then the numerator of $V_5'(r)$ is rewritten as
\bea
n_5&=&z^5 (-3 - 4 m (2 \mu^2 + (\sqrt{3} e - 4 \o) \o) - 4 \l_l)+
2 m z^4 (3 e^2 m - 12 m \mu^2 - 16 \sqrt{3} e m \o + 40 m \o^2 -8 \l_l)\nn\\
&&+ 2 m^2 z^3 (15 e^2 m - 12 m \mu^2 - 44 \sqrt{3} e m \o + 80 m \o^2 -10 \l_l)\nn\\
&& + 2 m^3 z^2 (27 e^2 m - 56 \sqrt{3} e m \o -4 (m (\mu^2 - 20 \o^2) + \l_l))\nn\\
&&+ 2 m^5 z(21 e^2 - 34 \sqrt{3} e \o + 40 \o^2) +4 m^6 (3 e^2 - 4 \sqrt{3} e \o + 4 \o^2)\nn\\
&&=\sum_{i=0}^5 a_i z^i,
\eea
where
\bea
a_5&=&-3 -8m\mu^2- 4 m \o (\sqrt{3} e - 4 \o) - 4 \l_l,\nn\\
a_4&=&2 m (3 e^2 m - 12 m \mu^2 - 16 \sqrt{3} e m \o + 40 m \o^2 -8 \l_l),\nn\\
a_3&=&2 m^2 (15 e^2 m - 12 m \mu^2 - 44 \sqrt{3} e m \o + 80 m \o^2 -10 \l_l),\nn\\
a_2&=&2 m^3(27 e^2 m - 56 \sqrt{3} e m \o -4 (m (\mu^2 - 20 \o^2) + \l_l)),\nn\\
a_1&=& 2 m^5(21 e^2 - 34 \sqrt{3} e \o + 40 \o^2),\nn\\
a_0&=&4 m^6 (3 e^2 - 4 \sqrt{3} e \o + 4 \o^2)=4m^6(2\o-\sqrt{3}e)^2.
\eea
A real root of $n_5(r)=0 $ when $r>r_h^{(5)}$ corresponds to a positive root of $n_5(z)=0$ when $z>0$.
In the next, we will prove there is indeed only one positive real root for the equation $n_5(z)=0$ by analyzing the signs of coefficients of $n_5(z)$.

It is obvious that $a_0>0$. Given the bound state condition and superradiance condition, $\o^2<\mu^2,\o<\frac{\sqrt{3}}{2} e$, it is easy to prove that $a_5<0$,
\bea
a_5&=&-3 -8m\mu^2- 4 m \o (\sqrt{3} e - 4 \o) - 4 \l_l\nn\\
&=&-3-8m(\mu^2-\o^2)-4m\o(\sqrt{3} e - 2 \o)-4\l_l<0.
\eea
The other coefficients can be rewritten as following
\bea
a_4&=&2 m (3 e^2 m - 12 m \mu^2 - 16 \sqrt{3} e m \o + 40 m \o^2 -8 \l_l)\nn\\
&=&2m(-8 \l_l+12m\o^2-12m\mu^2)+2m^2e^2(3- 16 \sqrt{3}t+28t^2),\\
a_3&=&2 m^2 (15 e^2 m - 12 m \mu^2 - 44 \sqrt{3} e m \o + 80 m \o^2 -10 \l_l)\nn\\
&=&2m^2(-10 \l_l+12m\o^2- 12 m \mu^2)+2m^3e^2(15- 44 \sqrt{3}t+68t^2),\\\label{a2}
a_2&=&2 m^3(27 e^2 m - 56 \sqrt{3} e m \o -4 (m (\mu^2 - 20 \o^2) + \l_l))\nn\\
&=&2 m^3(-4\l_l-4m\mu^2+4m\o^2)+2m^4e^2(27-56 \sqrt{3}t+76t^2),\\
a_1&=& 2 m^5(21 e^2 - 34 \sqrt{3} e \o + 40 \o^2)\nn\\
&=&2m^5e^2(21-34 \sqrt{3}t+40t^2),
\eea
where $t=\o/e$. For $a_1$, one can check $a_1<0$ is equivalent to
\bea\label{a11}
\frac{7\sqrt{3}}{20}<t<\frac{\sqrt{3}}{2},~or~0.61<t<0.87.
\eea
For $a_2$, the first term in \eqref{a2} is negative, so a sufficient condition for $a_2<0$ is that the second term in \eqref{a2} is also negative, i.e.
\bea\label{a22}
\frac{9\sqrt{3}}{38}<t<\frac{\sqrt{3}}{2},~or~0.41<t<0.87.
\eea
Using similar analysis, a sufficient condition for $a_3<0$  is
\bea\label{a33}
\frac{5\sqrt{3}}{34}<t<\frac{\sqrt{3}}{2},~or~0.25<t<0.87.
\eea
A sufficient condition for $a_4<0$  is
\bea\label{a44}
\frac{\sqrt{3}}{14}<t<\frac{\sqrt{3}}{2},~or~0.12<t<0.87.
\eea

Let us analyse the signs of coefficients $\{a_5,a_4,a_3,a_2,a_1,a_0\}$ with the varying of the parameter $t$ from 0 to $\frac{\sqrt{3}}{2}$.

When $0.61<t<0.87$, according to the equations \eqref{a11} to \eqref{a44}, all $a_i (i=1,2,3,4)$ are negative, the signs of the six coefficients $\{a_5,a_4,a_3,a_2,a_1,a_0\}$ are $\{-----+\}$.

When $0.41<t<0.61$, $a_1$ is positive and all other coefficients are negative. The signs of the six coefficients are $\{----++\}$.

When $0.25<t<0.41$, the sign of $a_2$ is not fixed by the above analysis and the signs of the six coefficients may be $\{----++\}$ or $\{---+++\}$.

For further analysis, we consider following two  differences,
\bea
\frac{a_2}{8m^3}-\frac{a_3}{20m^2}&=&\frac{ me^2}{20} (105- 192 \sqrt{3}t + 240 t^2+ 4 \mu^2/e^2),\\
\frac{a_3}{20m^2}-\frac{a_4}{16m}&=&\frac{3me^2}{40} (15- 32 \sqrt{3}t + 40 t^2 + 4 \mu^2/e^2 ).
\eea
One can check that
\bea\label{a2a3}
\frac{a_2}{8m^3}>\frac{a_3}{20m^2}
\eea
for $0<t<0.25$,  and
\bea\label{a3a4}
\frac{a_3}{20m^2}>\frac{a_4}{16m}
\eea
for $0<t<0.12$.
This implies if $a_3>0$, then $a_2>0$ for $0<t<0.25$ and if $a_4>0$, then $a_2, a_3 >0$ for $0<t<0.12$.

When $0.12<t<0.25$, according to the equations \eqref{a11} to \eqref{a44}, $a_4<0$ and $a_1>0$. The signs of $a_2, a_3$ are not fixed. But given equation \eqref{a2a3}, the possible signs of $\{a_3, a_2\}$ are
$\{-,-\}, \{-,+\},\{+,+\}$. Then the signs of the six coefficients may be $\{----++\}$, $\{---+++\}$, or $\{--++++\}$.

When $0<t<0.12$, we can make a similar analysis as the above case and find that the signs of the six coefficients may be $\{----++\}$, $\{---+++\}$, $\{--++++\}$, or $\{-+++++\}$.

Based on the above analysis on the possible signs of coefficients $\{a_5,a_4,a_3,a_2,a_1,a_0\}$, we conclude that
the number of sign changes in the sequence of the six coefficients is always 1 for $0<t<\frac{\sqrt{3}}{2}$. According to
Descartes' rule of signs, the polynomial equation $n_5(z)=0$ has at most one positive real root, which means the effective potential has at most one extreme outside the horizon. And we already know that there is at least one extreme (maximum) for the effective potential outside the horizon based on  the asymptotic analysis of $V_5(r)$. So there is  only one maximum for the effective potential  $V_5(r)$ outside the horizon and no potential well exists. The D=5 extremal RN black hole is superradiantly stable.

\section{D=6 extremal RN black holes}
For a D=6 extremal RN black hole, the event horizon is $r^{(6)}_h=m^{1/3}$.
The explicit expression of superradiance condition \eqref{sup-con-extr} is
\bea\label{sup-con-6D}
\o<\sqrt{\frac{{3}}{2}} e.
\eea
The denominator of the derivative of effective potential $V_6'(r)$ is $2 r^3(r^3-m)^5$. The numerator of $V_6'(r)$ is
\bea\nn
n_6=-4(\l_l+2)r^{15}+4 (-3 m \mu^2 - \sqrt{6} e m \o + 6 m \o^2)r^{14}+40m r^{12}\\\nn
+4 (2 e^2 m^2 + 6 m^2 \mu^2 - 3 \sqrt{6} e m^2 \o) r^{11}+4 m^2(-20 + 3 \l_l)r^9\\
+4 m^3(2 e^2 - 3 \mu^2)r^8+8 m^3(10 -\l_l)r^6- 40 m^4 r^3+8 m^5.
\eea
One can check that when $r$ goes to infinity, the asymptotic behavior of $V_6'(r)$ is
\bea\label{6Asymp}
V_6'\rightarrow \frac{-2(\l_l+2)}{r^3}+{\cal O}( \frac{1}{r^4}),
\eea
which is negative. Thus, there is no trapping potential well near the spatial infinity. This result is consistent with the general discussion before.

Now we change the radial variable from $r$ to $z=r-r^{(6)}_h=r-m^{1/3}$. The numerator of the derivative of the effective potential, $n_6$, is rewritten as
\bea\label{n6z}
&&n^{(6)}(z)=(-8 - 4 \l_l) z^{15}+4(-30 m^{1/3} - 3 m \mu^2 - \sqrt{6} e m \o + 6 m \o^2 -
 15 m^{1/3} \l_l)z^{14}\nn\\
&+&4(-210 m^{2/3} - 42 m^{4/3} \mu^2 - 14 \sqrt{6} e m^{4/3} \o +
 84 m^{4/3} \o^2 - 105 m^{2/3} \l_l)z^{13} \nn\\
 &+&4(-900 m - 273 m^{5/3} \mu^2 - 91 \sqrt{6} e m^{5/3} \o +
 546 m^{5/3} \o^2 - 455 m \l_l)z^{12}\nn\\
  &+&4(-2610 m^{4/3} + 2 e^2 m^2 - 1086 m^2 \mu^2 - 367 \sqrt{6} e m^2 \o +
 2184 m^2 \o^2 - 1365 m^{4/3} \l_l)z^{11}\nn\\
 &+&4(-5346 m^{5/3} + 22 e^2 m^{7/3} - 2937 m^{7/3} \mu^2 -
 1034 \sqrt{6} e m^{7/3} \o + 6006 m^{7/3} \o^2 - 3003 m^{5/3} \l_l)z^{10}\nn\\
 &+&4(-7830 m^2 + 110 e^2 m^{8/3} - 5676 m^{8/3} \mu^2 -
 2167 \sqrt{6} e m^{8/3} \o + 12012 m^{8/3} \o^2 - 5002 m^2 \l_l)z^{9}\nn\\
  &+&4(-8100 m^{7/3} + 332 e^2 m^3 - 8022 m^3 \mu^2 - 3498 \sqrt{6} e m^3 \o +
 18018 m^3 \o^2 - 6408 m^{7/3} \l_l)z^{8}\nn\\
  &+&4(-5670 m^{8/3} + 676 e^2 m^{10/3} - 8340 m^{10/3} \mu^2 -
 4422 \sqrt{6}e m^{10/3} \o + 20592 m^{10/3} \o^2 -
 6327 m^{8/3} \l_l)z^{7}\nn\\
 & +&4(-2430 m^3 + 980 e^2 m^{11/3} - 6321 m^{11/3} \mu^2 -
 4389 \sqrt{6} e m^{11/3} \o + 18018 m^{11/3} \o^2 - 4755 m^3 \l_l)z^{6}\nn\\
  &+&4(-486 m^{10/3} + 1036 e^2 m^4 - 3402 m^4 \mu^2 - 3388 \sqrt{6} e m^4 \o +
 12012 m^4 \o^2 - 2637 m^{10/3} \l_l)z^{5}\nn\\
 & +&4(800 e^2 m^{13/3} - 1233 m^{13/3} \mu^2 - 1991 \sqrt{6} e m^{13/3} \o +
 6006 m^{13/3} \o^2 - 1017 m^{11/3} \l_l)z^{4}\nn\\
 & +&4(442 e^2 m^{14/3} - 270 m^{14/3} \mu^2 - 859 \sqrt{6} e m^{14/3} \o +
 2184 m^{14/3} \o^2 - 243 m^4 \l_l)z^{3}\nn\\
 &+&4(166 e^2 m^5 - 27 m^5 \mu^2 - 256 \sqrt{6} e m^5 \o + 546 m^5 \o^2 -
 27 m^{13/3} \l_l)z^{2}\nn\\
 & +&4(38 e^2 m^{16/3} - 47 \sqrt{6} e m^{16/3} \o + 84 m^{16/3} \o^2)z+8 m^{17/3}(\sqrt{2}e-\sqrt{3}\o)^2\\
 &\equiv& \sum_{i=0}^{15} b_i z^i.
\eea

We will prove that the polynomial equation $n_6(z)=0$ has at most one positive real root in the following.
According to Descartes' rule of signs, we need to prove the number of sign changes is 1 in the sequence of the polynomial's sixteen coefficients, $\{b_{15}, b_{14}, ..., b_0\}$.

It is easy to see that
\bea\label{b0}
b_0=8 m^{17/3}(\sqrt{2}e-\sqrt{3}\o)^2>0,~~b_{15}=-8 - 4 \l_l<0.
\eea
Let's check the sign of $b_{14}$. We rewrite $b_{14}$ as following
\bea
b_{14}&=&4(-30 m^{1/3} - 3 m \mu^2 - \sqrt{6} e m \o + 6 m \o^2 -
 15 m^{1/3} \l_l)\nn\\
 &=&4[-30 m^{1/3} - 15 m^{1/3} \l_l+(- 3 m \mu^2+3m\o^2)+(3m\o^2 - \sqrt{6} e m \o)].
\eea
Taking into account the bound state and superradiance conditions, $\o<\mu,~\o<e\Phi_h=\sqrt{2/3}e$, the two terms in parentheses of the above equation are negative and then
\bea
b_{14}<0.
\eea
Similarly, we can easily prove
\bea
b_{13}<0,~~b_{12}<0.
\eea
However, it is not easy to prove the signs of $b_1, b_2,.., b_{11}$ similarly.

Let's study the relation between the signs of $b_1$ and $ b_2$. These two coefficients can be read out directly from \eqref{n6z},
\bea
b_1&=&4(38 e^2 m^{16/3} - 47 \sqrt{6} e m^{16/3} \o + 84 m^{16/3} \o^2),\\
b_2&=&4(166 e^2 m^5 - 27 m^5 \mu^2 - 256 \sqrt{6} e m^5 \o + 546 m^5 \o^2 -
 27 m^{13/3} \l_l).
\eea
Define two new coefficients $b'_1, b'_2$,
\bea
b'_1=\frac{b_1}{4*84 m^{16/3}},~b'_2=\frac{b_2}{4*519 m^5}.
\eea
The difference between $b'_1$ and $ b'_2$ is
\bea
b'_1-b'_2=\frac{321 e^2}{2422}+\frac{256}{173} \sqrt{\frac{2}{3}} e \o-\frac{47 e \o}{14 \sqrt{6}}+\frac{9 \lambda_l }{173 m^{2/3}}+\frac{9 }{173}({\mu}^2- \o^2).
\eea
Given the bound state condition, $\o^2<\mu^2$, the last term in the above equation is positive.
Given the superradiance condition \eqref{sup-con-6D}, one can check the sum of the first three terms in the above equation is positive.
It is obvious that the $\l_l$ term in the above equation is also positive. So we have
\bea
b'_1-b'_2>0.
\eea
The possible signs of $\{b_2,b_1\}$ is $\{-,-\},\{-,+\},\{+,+\}$. The sign of $b_1$ is not smaller than the sign of $b_2$, i.e.
\bea\label{b1b2}
\text{sign}(b_2)\leqslant \text{sign}(b_1).
\eea

Next, let's study the relation between the signs of $b_2$ and $ b_3$. The coefficient $b_3$ can be read out directly from \eqref{n6z},
\bea
b_3=4(442 e^2 m^{14/3} - 270 m^{14/3} \mu^2 - 859 \sqrt{6} e m^{14/3} \o +
 2184 m^{14/3} \o^2 - 243 m^4 \l_l).
\eea
Define a new coefficient $b'_3$,
\bea
b'_3=\frac{b_3}{7656 m^{14/3}}.
\eea
The difference between $b'_2$ and $b'_3$ is
\bea
b'_2-b'_3=\frac{4907 e^2}{55187}-\frac{256}{173} \sqrt{\frac{2}{3}} e \o+\frac{859 e \o}{319 \sqrt{6}}+\frac{8271 \lambda_l }{110374 m^{2/3}}+\frac{4914}{55187}( {\mu}^2-\o^2).
\eea
Given the bound state condition, $\o^2<\mu^2$, the last term in the above equation is positive.
Given the superradiance condition \eqref{sup-con-6D}, one can check the sum of the first three terms in the above equation is positive.
It is obvious that the $\l_l$ term in the above equation is also positive. So we have
\bea
b'_2-b'_3>0.
\eea
The possible signs of $\{b_3,b_2\}$ is $\{-,-\},\{-,+\},\{+,+\}$. The sign of $b_2$ is not smaller than the sign of $b_3$, i.e.
\bea\label{b2b3}
\text{sign}(b_3)\leqslant \text{sign}(b_2).
\eea

In order to study the relation between the signs of $b_3$ and $ b_4$. Define a new parameter
\bea
b'_4=\frac{b_4}{4*4473 m^{13/3}}.
\eea
The difference between $b'_3$ and $b'_4$ is
\bea
b'_3-b'_4=\frac{32137 e^2}{507529}+\frac{1991 \sqrt{\frac{2}{3}} e \o}{1591}-\frac{859 e \o}{319 \sqrt{6}}+\frac{87411 \lambda_l }{1015058 m^{2/3}}+\frac{59514 }{507529}(\mu^2-\o^2).
\eea
We can prove the above difference is positive with the same method as previous cases. So the possible signs of $\{b_4,b_3\}$ is $\{-,-\},\{-,+\},\{+,+\}$, i.e.
\bea\label{b3b4}
\text{sign}(b_4)\leqslant \text{sign}(b_3).
\eea

Similarly, define the following new coefficients
\bea
b'_5=\frac{b_5}{4*8610 m^4}, b'_6=\frac{b_6}{4*11697 m^{11/3}}, b'_7=\frac{b_7}{4*12252 m^{10/3}},\\ b'_8=\frac{b_8}{39984 m^3}, b'_9=\frac{b_9}{25344 m^{8/3}},
b'_{10}=\frac{b_{10}}{12276 m^{7/3}}, b'_{11}=\frac{b_{11}}{4392 m^2}.
\eea
Given the bound state condition, $\o^2<\mu^2$, and the superradiance condition \eqref{sup-con-6D}, we find that [see the Appendix]
\bea
b'_4>b'_5>b'_6>b'_7>b'_8>b'_9>b'_{10}>b'_{11}.
\eea
So we have
\bea\label{b11}
\text{sign}(b_{11})\leqslant \text{sign}(b_{10})\leqslant \text{sign}(b_9)\leqslant \text{sign}(b_8)
\leqslant \text{sign}(b_7) \leqslant \text{sign}(b_6)\leqslant \text{sign}(b_5)\leqslant \text{sign}(b_4).
\eea

With the results  \eqref{b1b2}, \eqref{b2b3}, \eqref{b3b4},\eqref{b11}, the possible signs of $\{b_{11},...,b_1\}$ may be
all plus, $\{+,+,..,+,+\}$, all minus $\{-,-,..,-,-\}$, or $\{-,...,-,+,...,+\}$.
The signs of the sixteen coefficients of $n_6(z)$, $\{b_{15},b_{14},b_{13},...,b_{1},b_0\}$, are
$\{-,-,-,-,*,*,...,*,+\}$. The $*$ parts are the signs of $\{b_{11},...,b_1\}$. It is obvious that for all possible signs of $\{b_{11},...,b_1\}$, the number of the sign changes in the sequence of the sixteen coefficients is always 1.

So there is at most one extreme for the effective potential $V_6(r)$ outside the horizon and together with our
previous asymptotic analysis of $V_6(r)$, we conclude that there is only one maximum for the effective potential $V_6(r)$ outside the horizon and no potential well exists. The D=6 extremal RN black hole is superradiantly stable.

\section{Conclusion and discussion}
In this paper, superradiant stability of  D=5,6 extremal RN black holes  under charged massive scalar perturbation is investigated. A new method is developed that depends mainly on Descartes' rule of signs for the polynomial equations. In $D$=5 case, based on the asymptotic analysis of the effective potential $V(r)$ \eqref{asymp}, we know there is at least one extreme for the effective potential outside the horizon. With the new method, we prove that the derivative of the effective potential has at most one extreme outside the horizon \eqref{table}. There is only one maximum for the effective potential outside the horizon and no potential well exists, so the extremal RN black hole is superradiantly stable.  In $D$=6 case, the asymptotic analysis \eqref{6Asymp} shows that the effective potential has at least one extreme outside the horizon. According to the sign relations  \eqref{b1b2}, \eqref{b2b3}, \eqref{b3b4},\eqref{b11} and Descartes' rule of signs, we know there is at most one extreme for the effective potential outside the horizon. There is only one maximum for the effective potential outside the horizon and no potential well exists, so the six-dimensional extremal RN black hole is also superradiantly stable.

In $D$=6 case, it is quite unexpected that there are such interesting sign relations \eqref{b1b2}, \eqref{b2b3}, \eqref{b3b4},\eqref{b11}  between the signs of  the complicated coefficients in $n_6(z)$. It does not seem to be a coincidence. Due to this observation, we also have finished part of the proof for $D$=7 case and found  similar interesting sign relations. So we conjecture that  all $D$-dimensional ($D\geqslant 5$) extremal RN black holes are superradiantly stable under charged massive scalar perturbation which is minimally coupled with the black holes. It will be interesting to give a general proof for this, although it is not an easy work.

\begin{acknowledgements}

This work is partially supported by Guangdong Major Project of Basic and Applied Basic Research (No. 2020B0301030008), Science and Technology Program of Guangzhou (No. 2019050001) and Natural Science Foundation of Guangdong Province (No. 2020A1515010388).

\end{acknowledgements}

\appendix*
\section{}
In this appendix, we will present the mentioned details in the proof of $D$=6 case. Define the following scaled coefficients
\bea
b'_5=\frac{b_5}{4*8610 m^4}, b'_6=\frac{b_6}{4*11697 m^{11/3}}, b'_7=\frac{b_7}{4*12252 m^{10/3}},\\ b'_8=\frac{b_8}{39984 m^3}, b'_9=\frac{b_9}{25344 m^{8/3}},
b'_{10}=\frac{b_{10}}{12276 m^{7/3}}, b'_{11}=\frac{b_{11}}{4392 m^2}.
\eea
We have the following differences
\bea
b'_4-b'_5=\frac{15422 e^2}{326155}-\frac{7711 \sqrt{6} e \o}{326155}+\frac{425559 \lambda_l }{4566170 m^{2/3}}\nn\\+\frac{81}{1435 m^{2/3}}+\frac{44616}{326155}(\mu^2-\o^2),
\eea
\bea
b'_5-b'_6=\frac{12518 e^2}{342555}-\frac{242}{205} \sqrt{\frac{2}{3}} e\o+\frac{209}{557} \sqrt{6} e \o+\frac{160247 \lambda_l }{1598590 m^{2/3}}\nn\\+\frac{120933}{799295 m^{2/3}}+\frac{16588 }{114185}(\mu^2-\o^2),
\eea

\bea
b'_6-b'_7=\frac{16269 e^2}{568697}-\frac{209}{557} \sqrt{6} e \o+\frac{737 \sqrt{\frac{3}{2}} e \o}{1021}+\frac{1749851 \lambda_l }{15923516 m^{2/3}}\nn\\+\frac{2030535}{7961758 m^{2/3}}+\frac{79794 }{568697}(\mu^2-\o^2),
\eea

\bea
b'_7-b'_8=\frac{18678 e^2}{850493}-\frac{9339 \sqrt{6} e \o}{850493}+\frac{424059 \lambda_l }{3401972 m^{2/3}}\nn\\+\frac{591165}{1700986 m^{2/3}}+\frac{29601}{242998}(\mu^2-\o^2),
\eea

\bea
b'_8-b'_9=\frac{3803 e^2}{239904}-\frac{583}{833} \sqrt{\frac{3}{2}} e \o+\frac{197 e \o}{96 \sqrt{6}}+\frac{391621 \lambda_l }{2638944 m^{2/3}}\nn\\+\frac{124755}{293216 m^{2/3}}+\frac{533 }{5712}(\mu^2-\o^2)
\eea
\bea
b'_9-b'_{10}=\frac{91 e^2}{8928}+\frac{94}{93} \sqrt{\frac{2}{3}} e \o-\frac{197 e \o}{96 \sqrt{6}}+\frac{18565 \lambda_l }{98208 m^{2/3}}\nn\\+\frac{5523}{10912 m^{2/3}}+\frac{91}{1488}(\mu^2-\o^2)
\eea
\bea
b'_{10}-b'_{11}=\frac{91 e^2}{17019}-\frac{94}{93} \sqrt{\frac{2}{3}} e \o+\frac{367 e \o}{183 \sqrt{6}}+\frac{1001 \lambda_l }{3782 m^{2/3}}\nn\\+\frac{1201}{1891 m^{2/3}}+\frac{182}{5673}(\mu^2-\o^2).
\eea
One can easily check that all the differences are positive under bound state and superradiance conditions.

\end{document}